\documentclass[12pt]{article}

\usepackage{epsfig}
\usepackage{tabularx}
\usepackage{graphicx}
\usepackage{amssymb,amsfonts,amsmath}

\newcommand{\mchi}{m_{\chi}}
\newcommand{\mgravitino}{m_{\gravitino}}

\newcommand{\gravitino}{\tilde{G}}

\newcommand{\rem}[1]{{}}

\begin{document}

\begin{titlepage}

\centerline{\large \bf {Gravitino dark matter in the constrained}}
\centerline{\large \bf {next-to-minimal supersymmetric standard model}}
\centerline{\large \bf {with neutralino next-to-lightest superpartner}}

\vskip 1cm

\centerline{Gabriela Barenboim and Grigoris Panotopoulos}

\vskip 1cm

\centerline{Departamento de Fisica Teorica, University of Valencia and IFIC}

\vskip 0.2 cm

\centerline{Av. Moliner 50, 46100 Valencia, Spain}

\vskip 0.2 cm

\centerline{gabriela.barenboim@uv.es, grigoris.panotopoulos@uv.es}

\begin{abstract}
The viability of a possible cosmological scenario is investigated.
The theoretical framework is the constrained next-to-minimal
supersymmetric standard model (cNMSSM), with a gravitino playing the
role of the lightest
supersymmetric particle (LSP) and a neutralino acting as the
next-to-lightest
supersymmetric particle (NLSP). All the necessary constraints from
colliders and cosmology have been taken into account. For gravitino
we have considered the two usual production mechanisms, namely
out-of equillibrium decay from the NLSP, and scattering processes
from the thermal bath. The maximum allowed reheating temperature
after inflation, as well as the maximum allowed gravitino mass are
determined.
\end{abstract}

\end{titlepage}


\section{Introduction}

There is accumulated evidence both from astrophysics and cosmology
that about 1/4 of the energy budget of the universe consists of so
called dark matter, namely a component which is non-relativistic and
neither feels the electromagnetic nor the strong interaction. For a
review on dark matter see e.g.~\cite{Munoz:2003gx}. Although the
list of possible dark matter candidates is long, it is fair to say
that the most popular dark matter candidate is the lightest
supersymmetric particle (LSP) in supersymmetric models with R-parity
conservation~\cite{Feng:2003zu}. The superpartners that have the
right properties for playing the role of cold dark matter in the
universe are the axino, the gravitino and the lightest neutralino.
By far the most discussed case in the literature is the case of the
neutralino (see the classic review ~\cite{Jungman:1995df}), probably
because of the prospects of possible detection.

However, the gravitino is another very interesting candidate for
cold dark matter, since its interactions are completely determined
by the supergravity lagrangian~\cite{Cremmer:1982en}, in contrast to
what happens to the neutralino or the axino case, where the
interactions depend on the chosen model. Unfortunately, gravitino
belongs to the class of new exotic particles that can be potentially
dangerous for cosmology, and it is therefore escorted by the
so-called gravitino problem~\cite{Ellis:1982yb}. The mass of the
gravitino strongly depends on the SUSY-breaking scheme, and can
range from eV scale to scales beyond the TeV
region~\cite{Giudice:1998bp, Nilles:1983ge, Randall:1998uk}. In
particular, in gauge-mediated SUSY-breaking
schemes~\cite{Giudice:1998bp} the gravitino mass is typically less
than 100~MeV, while in gravity-mediated schemes~\cite{Nilles:1983ge}
it is expected to be in the GeV to TeV range. Finally, it must be
noted that there are hybrid models of gauge- and gravity-mediation,
in which gravity provides sub-dominant and yet non-negligible
contributions~\cite{Hiller:2008sv}. Therefore, according to the
precise mechanism for supersymmetry breaking, the gravitino can be
either stable or unstable, with the corresponding gravitino
cosmology. In general, the gravitino problem requires that the
reheating temperature after inflation should be lower than
$10^6-10^7$~GeV~\cite{Kawasaki:2004yh, Pradler:2006hh}, which poses
serious difficulties to the thermal leptogenesis
scenario~\cite{Buchmuller:2002rq}.

In the present work we want to consider gravitino dark matter in the
constrained next-to-minimal supersymmetric standard model (cNMSSM),
assuming that the lightest neutralino is the next-to-lightest
supersymmetric particle (NLSP), and taking into account the two
usual gravitino production mechanisms to be discussed later on,
namely the out-of-equillibrium decays of the NLSP, as well as
scattering processes from the thermal bath.

This article is organized as follows. In the next section we present
the theoretical framework. In section 3 we discuss all the relevant
constraints from colliders and from cosmology, and we show our
results. Finally, we conclude.

\section{Theoretical framework}

In the present article we work in the framework of the constrained
next-to-minimal supersymmetric standard model (cNMSSM). We assume
that the gravitino is the LSP, while the lightest neutralino is the
NLSP. The gravitino is stable and plays the role of cold dark matter
in the universe, while the neutralino is unstable and it decays to
gravitino.

In what folows we review in short the particle physics model, namely
the cNMSSM, as well as the gravitino production mechanisms.

\subsection{Basics of cNMSSM}

The most straightforward extension of standard model (SM) of
particle physics based on SUSY is the minimal supersymmetric
standard model (MSSM)~\cite{mssm}. It is a supersymmetric gauge
theory based on the SM gauge group with the usual representations
(singlets, doublets, triplets) and on $\mathcal{N}=1$ SUSY.
Excluding gravity, the massless representations of the SUSY algebra
are a chiral and a vector supermultiplet. The gauge bosons and the
gauginos are members of the vector supermultiplet, while the matter
fields (quarks, leptons, Higgs) and their superpartners are members
of the chiral supermultiplet. The Higgs sector in the MSSM is
enhanced compared to the SM case. There are now two Higgs doublets,
$H_u, H_d$, (or $H_1, H_2$) for anomaly cancelation requirements and
for giving masses to both up and down quarks. After electroweak
symmetry breaking we are left with five physical Higgs bosons, two
charged $H^{\pm}$ and three neutral $A,H,h$ ($h$ being the
lightest). Since we have not seen any superpartners yet, SUSY has to
be broken. In MSSM, SUSY is softly broken by adding to the
Lagrangian terms of the form
\begin{itemize}
\item Mass terms for the gauginos $\tilde{g}_i$, $M_1, M_2, M_3$
\begin{equation}
M \tilde{g} \tilde{g}
\end{equation}
\item Mass terms for sfermions $\tilde{f}$
\begin{equation}
m_{\tilde{f}}^2 \tilde{f}^{\dag} \tilde{f}
\end{equation}
\item Masses and bilinear terms for the Higgs bosons $H_u, H_d$
\begin{equation}
m_{H_u}^2 H_u^{\dag} H_u+m_{H_d}^2 H_d^{\dag} H_d+B \mu (H_u H_d +
h.c.)
\end{equation}
\item Trilinear couplings between sfermions and Higgs bosons
\begin{equation}
A Y \tilde{f}_1 H \tilde{f}_2
\end{equation}
\end{itemize}
In the unconstrained MSSM there is a huge number of unknown
parameters~\cite{parameters} and thus little predictive power.
However, motivated by the grand unification idea, the constrained
MSSM (CMSSM) assumes that gaugino masses, scalar masses and
trilinear couplings have (separately) a common, universal, value at
the GUT scale, like the gauge coupling constants do. CMSSM is
therefore a framework with a small controllable number of
parameters, and thus with much more predictive power. In the CMSSM
there are four parameters, $m_0, m_{1/2}, A_0, tan \beta$, which are
explained below, plus the sign of the $\mu$ parameter from the Higgs
sector. The magnitude of $\mu$, as well as the B parameter mentioned
above, are determined by the requirement for a proper electroweak
symmetry breaking. However, the sign of $\mu$ remains undetermined.
The other four parameters of the CMSSM are related to
\begin{itemize}
\item Universal gaugino masses
\begin{equation}
M_1(M_{GUT})=M_2(M_{GUT})=M_3(M_{GUT})=m_{1/2}
\end{equation}
\item Universal scalar masses
\begin{equation}
m_{\tilde{f}_i}(M_{GUT})=m_0
\end{equation}
\item Universal trilinear couplings
\begin{equation}
A_{i j}^u(M_{GUT}) = A_{i j}^d(M_{GUT}) = A_{i j}^l(M_{GUT}) = A_0 \delta_{i j}
\end{equation}
\item
\begin{equation}
tan \beta \equiv \frac{v_1}{v_2}
\end{equation}
where $v_1, v_2$ are the vevs of the Higgs doublets and $M_{GUT} \sim 10^{16}~GeV$ is
the Grand Unification
scale.
\end{itemize}

Unfortunately, the CMSSM suffers from the so-called $\mu$
problem~\cite{Kim:1983dt}. This problem is elegantly solved in the
framework of the next-to-minimal supersymmetric standard model
(NMSSM)~\cite{Nilles:1982dy}. In addition to the MSSM Yukawa
couplings for quarks and leptons, the NMSSM superpotential contains
two additional terms involving the Higgs doublet superfields, $H_1$
and $H_2$, and the new superfield $S$, a singlet under the SM gauge
group $SU(3)_c \times SU(2)_L \times U(1)_Y$~\cite{Cerdeno:2004xw}
\begin{equation}\label{2:Wnmssm}
W= \epsilon_{ij} \left( Y_u \, H_2^j\, Q^i \, u + Y_d \, H_1^i\, Q^j
\, d + Y_e \, H_1^i\, L^j \, e \right) - \epsilon_{ij} \lambda \,S
\,H_1^i H_2^j +\frac{1}{3} \kappa S^3\,
\end{equation}
where we take $H_1^T=(H_1^0, H_1^-)$, $H_2^T=(H_2^+, H_2^0)$, $i,j$ are
$SU(2)$ indices, and $\epsilon_{12}=1$. In this model, the usual MSSM bilinear
$\mu$ term is absent from the superpotential, and only dimensionless trilinear
couplings are present in $W$. However, when the scalar component of $S$ acquires
a VEV, an effective interaction $\mu H_1 H_2$ is generated, with $\mu \equiv
\lambda \langle S \rangle$.

Finally, the soft SUSY
breaking terms are given by~\cite{Cerdeno:2004xw}
\begin{align}\label{2:Vsoft}
-\mathcal{L}_{\text{soft}}=&\,
 {m^2_{\tilde{Q}}} \, \tilde{Q}^* \, \tilde{Q}
+{m^2_{\tilde{U}}} \, \tilde{u}^* \, \tilde{u}
+{m^2_{\tilde{D}}} \, \tilde{d}^* \, \tilde{d}
+{m^2_{\tilde{L}}} \, \tilde{L}^* \, \tilde{L}
+{m^2_{\tilde{E}}} \, \tilde{e}^* \, \tilde{e}
 \nonumber \\
&
+m_{H_1}^2 \,H_1^*\,H_1 + m_{H_2}^2 \,H_2^* H_2 +
m_{S}^2 \,S^* S \nonumber \\
&
+\epsilon_{ij}\, \left(
A_u \, Y_u \, H_2^j \, \tilde{Q}^i \, \tilde{u} +
A_d \, Y_d \, H_1^i \, \tilde{Q}^j \, \tilde{d} +
A_e \, Y_e \, H_1^i \, \tilde{L}^j \, \tilde{e} + \text{H.c.}
\right) \nonumber \\
&
+ \left( -\epsilon_{ij} \lambda\, A_\lambda S H_1^i H_2^j +
\frac{1}{3} \kappa \,A_\kappa\,S^3 + \text{H.c.} \right)\nonumber \\
& - \frac{1}{2}\, \left(M_3\, \lambda_3\, \lambda_3+M_2\,
\lambda_2\, \lambda_2 +M_1\, \lambda_1 \, \lambda_1 + \text{H.c.}
\right) \,
\end{align}

Clearly, the NMSSM is very similar to the MSSM. Despite the similarities
between the two particle physics models, the Higgs sector as well as the neutralino
mass matrix and mass eigenstates in the NMSSM are more complicated
compared to the corresponding ones in the MSSM.

In particular, in the Higgs sector we have now two CP-odd neutral,
and three CP-even neutral Higgses. We make the assumption that there
is no CP-violation in the Higgs sector at tree level, and neglecting 
loop level effects the CP-even
and CP-odd states do not mix. We are not interested in the CP-odd
states, while the CP-even Higgs interaction and physical eigenstates
are related by the transformation
\begin{equation}\label{2:Smatrix}
h_a^0 = S_{ab} H^0_b\,
\end{equation}
where $S$ is the unitary matrix that diagonalises the CP-even
symmetric mass matrix, $a,b = 1,2,3$, and the physical eigenstates
are ordered as $m_{h_1^0} \lesssim
m_{h_2^0} \lesssim m_{h_3^0}$.

In the neutralino sector the situation is again more involved, since
the fermionic component of $S$ mixes with the neutral Higgsinos,
giving rise to a fifth neutralino state. In the weak interaction
basis defined by ${\Psi^0}^T \equiv \left(\tilde B^0=-i
\lambda^\prime, \tilde W_3^0=-i \lambda_3, \tilde H_1^0, \tilde
H_2^0, \tilde S \right)\,$, the neutralino mass terms in the
Lagrangian are~\cite{Cerdeno:2004xw}
\begin{equation}
\mathcal{L}_{\mathrm{mass}}^{\tilde \chi^0} =
-\frac{1}{2} (\Psi^0)^T \mathcal{M}_{\tilde \chi^0} \Psi^0 + \mathrm{H.c.}\,,
\end{equation}
with $\mathcal{M}_{\tilde \chi^0}$ a $5 \times 5$ matrix,
{\footnotesize \begin{equation}
  \mathcal{M}_{\tilde \chi^0} = \left(
    \begin{array}{ccccc}
      M_1 & 0 & -M_Z \sin \theta_W \cos \beta &
      M_Z \sin \theta_W \sin \beta & 0 \\
      0 & M_2 & M_Z \cos \theta_W \cos \beta &
      -M_Z \cos \theta_W \sin \beta & 0 \\
      -M_Z \sin \theta_W \cos \beta &
      M_Z \cos \theta_W \cos \beta &
      0 & -\lambda s & -\lambda v_2 \\
      M_Z \sin \theta_W \sin \beta &
      -M_Z \cos \theta_W \sin \beta &
      -\lambda s &0 & -\lambda v_1 \\
      0 & 0 & -\lambda v_2 & -\lambda v_1 & 2 \kappa s
    \end{array} \right)
  \label{neumatrix}
\end{equation}}
The above matrix can be diagonalised by means of a unitary matrix
$N$
\begin{equation}
N^* \mathcal{M}_{\tilde \chi^0} N^{-1} = \operatorname{diag}
(m_{\tilde \chi^0_1}, m_{\tilde \chi^0_2}, m_{\tilde \chi^0_3},
m_{\tilde \chi^0_4}, m_{\tilde \chi^0_5})\,
\end{equation}
where $m_{\tilde \chi^0_1}$ is the lightest
neutralino mass. Under the above assumptions, the lightest neutralino can be
expressed as the combination
\begin{equation} \label{composition}
\tilde \chi^0_1 = N_{11} \tilde B^0 + N_{12} \tilde W_3^0 + N_{13}
\tilde H_1^0 + N_{14} \tilde H_2^0 + N_{15} \tilde S\,
\end{equation}
In the following, neutralinos with $N^2_{11}>0.9$, or
$N^2_{15}>0.9$, will be referred to as bino- or singlino-like,
respectively.

Similarly to the CMSSM, in the constrained next-to-minimal supersymmetric standard model
the universality of $m_0, A_0, m_{1/2}$ at the GUT scale is again assumed,
with the only parameters now being~\cite{Hugonie:2007vd} \\
\centerline{$tan \beta, m_0, A_0, m_{1/2}, \lambda, A_k$} and the
sign of the $\mu$ parameter can be chosen at will.

We end the discussion on the particle physics model here, by making
a final remark regarding the differences between the CMSSM and the
cNMSSM. In the CMSSM the lightest neutralino is mainly a bino in
most of the parameter space, and low values of $m_0$ are disfavored
because they lead to charged sleptons that are lighter than the
neutralino $\chi_1^0$, while in the cNMSSM the lightest neutralino
is mainly a singlino in large regions of the parameter space, thanks
to which the charged LSP problem can be
avoided~\cite{Hugonie:2007vd}. Furthermore, in the cNMSSM there are
more mechanisms that contribute to the neutralino relic
density~\cite{Hugonie:2007vd}.

\subsection{Gravitino production}

For the gravitino abundance we take the relevant production mechanisms into account
and impose the cold
dark matter constraint~\cite{Komatsu:2008hk}
\begin{equation}
0.1097 < \Omega_{cdm} h^2=\Omega_{3/2} h^2 < 0.1165
\end{equation}
At this point it is convenient to define the gravitino yield, $Y_{3/2} \equiv n_{3/2}/s$,
where $n_{3/2}$ is
the gravitino number density, $s=h_{eff}(T) \frac{2 \pi^2}{45} T^3$ is the entropy density
for a relativistic
thermal bath, and $h_{eff}$ counts the relativistic degrees of freedom. The gravitino
abundance $\Omega_{3/2}$
in terms of the gravitino yield is given by
\begin{equation}
\Omega_{3/2} h^2=\frac{\mgravitino s(T_0) Y_{3/2} h^2}{\rho_{cr}}=2.75 \times 10^{8} \left ( \frac{\mgravitino}{GeV}
\right ) Y_{3/2}(T_0)
\end{equation}
where we have used the values
\begin{eqnarray}
T_0 & = & 2.73 K=2.35 \times 10^{-13}~\textrm{GeV} \\
h_{eff}(T_0)& = & 3.91 \\
\rho_{cr}/h^2& = & 8.1 \times 10^{-47}~\textrm{GeV}^4
\end{eqnarray}
The total gravitino yield has two contributions, namely one from the thermal bath, and one
from the out-of-equillibrium NLSP decay.
\begin{equation}
Y_{3/2}=Y_{3/2}^{TP}+Y_{3/2}^{NLSP}
\end{equation}
The contribution from the thermal production has been computed in
~\cite{Bolz:2000fu,Pradler:2006qh,Rychkov:2007uq}. In~\cite{Bolz:2000fu} the gravitino
production was
computed in leading order in the gauge coupling $g_3$,
in~\cite{Pradler:2006qh} the thermal
rate was
computed in leading order in all Standard Model gauge couplings $g_Y, g_2, g_3$, and
in~\cite{Rychkov:2007uq}
new effects were taken into account, namely: a) gravitino production via gluon $\rightarrow$
gluino
$+$ gravitino and other decays, apart from the previously considered $2 \rightarrow 2$
gauge scatterings,
b) the effect of the top Yukawa coupling, and c) a proper treatment of the reheating
process.
Here we
shall use the result of~\cite{Bolz:2000fu} since the corrections
of~\cite{Pradler:2006qh,Rychkov:2007uq}
do not alter our conclusions. Therefore the thermal gravitino production is given by
\begin{equation}
Y_{3/2}^{TP}=0.29 \times 10^{-12} \: \left (
\frac{T_R}{10^{10}~GeV}\right ) \: \left (1+\frac{1}{3}
\frac{m_{\tilde{g}}^2}{\mgravitino^2}\right )
\end{equation}
or, approximately for a light gravitino, $\mgravitino \ll m_{\tilde{g}}$
\begin{equation}
Y_{3/2}^{TP} \simeq 10^{-13} \: \left ( \frac{T_R}{10^{10}~GeV}
\right ) \: \left ( \frac{m_{\tilde{g}}}{\mgravitino} \right )^2
\end{equation}
with $\mgravitino$ the gravitino mass and $m_{\tilde{g}}$ the gluino mass. At
this point it must be noted that all the relevant particles here (gluons,
quarks, gluinos, squarks
and gravitinos) are supposed to be in thermal equillibrium, and thus the reheating
temperature after inflation should be at least $1$~TeV. This is going to be
important later on.

The second contribution to the gravitino abundance comes from the decay of the NLSP
\begin{equation}
\Omega_{3/2}^{NLSP} h^2 = \frac{\mgravitino}{m_{NLSP}} \: \Omega_{NLSP} h^2
\end{equation}
with $m_{NLSP}$ the mass of the NLSP, and $\Omega_{NLSP} h^2$ the
abundance the NLSP would have, had it not decayed into the gravitino.

\section{Constraints and results}

- Spectrum and collider constraints: We have used
NMSSMTools~\cite{Ellwanger:2004xm}, a computer software that
computes the masses of the Higgses and the superpartners, the
couplings, and the relic density of the neutralino, for a given set
of the free parameters. We have performed a random scan in the whole
parameter space (with fixed $\mu > 0$ motivated by the muon
anomalous magnetic moment), and we have selected only those points
that satisfy i) theoretical requirements, such as neutralino LSP,
correct electroweak symmetry breaking, absence of tachyonic masses
etc., and ii) LEP bounds on the Higgs mass, collider bounds on SUSY
particle masses, and experimental data from
B-physics~\cite{precision, Yao:2006px}. For all these "good" points
the lightest neutralino is either a bino or a singlino, and contrary
to the case where neutralino is the dark matter particle, here we do
not require that the neutralino relic density falls within
the allowed WMAP range mentioned before.

- As we have already mentioned, the total gravitino abundance, and
not the neutralino one, should satisfy the cold dark matter
constraint~\cite{Komatsu:2008hk}
\begin{equation}
0.1097 < \Omega_{cdm} h^2=\Omega_{3/2} h^2 < 0.1165
\end{equation}
that relates the reheating temperature after inflation to the
gravitino mass as follows
\begin{equation}
0.11 = A(\mgravitino, m_{\tilde{g}}) T_R+\frac{m_{3/2}}{m_{NLSP}} \Omega_{NLSP} h^2
\end{equation}
For a given point in the cNMSSM parameter space, the complete
spectrum and couplings have been computed, and we are left with two
more free parameters, namely the gravitino mass and the reheating
temperature after inflation. The gravitino mass is directly related
to the SUSY-breaking scheme, while the precise range of values of
the reheating temperature is crucial for the baryon asymmetry
generation mechanism. The thermal production contribution cannot be
larger than the total gravitino abundance, and for this we can
already obtain an upper bound on the reheating temperature
\begin{equation}
T_R \leq 4.1 \times 10^9 \left ( \frac{\mgravitino}{100~GeV} \right ) \: \left
( \frac{TeV}{m_{\tilde{g}}} \right )^2 \: GeV
\end{equation}
Assuming a gluino mass $m_{\tilde{g}} \sim 1$~TeV, we can see that for a
heavy gravitino, $\mgravitino \sim 100$~GeV, it is possible to obtain a reheating
temperature large enough for thermal leptogenesis. However, for a light gravitino,
$\mgravitino \sim 1$~GeV, we cannot obtain a reheating temperature larger than
$T_R \sim 10^7$~GeV.

- In scenarios in which gravitino is assumed to be the LSP, the NSLP
is unstable with a lifetime that is typically larger than BBN time
$t_{BBN} \sim 1$~sec. Energetic particles produced by the NLSP decay
may dissociate the background nuclei and significantly affect the
primordial abundances of light elements. If such processes occur
with sizable rates, the predictions of the standard BBN scenario
would be altered and the success of the primordial nucleosynthesis
would be spoiled. BBN constraints on cosmological scenarios with
exotic long-lived particles predicted by physics beyond the Standard
Model have been studied~\cite{Kawasaki:2004yh, Cyburt:2002uv}.
Previous investigations have shown that the neutralino NLSP scenario
with a gravitino mass $m_{3/2} \geq 100$~MeV and a neutralino
lifetime in the range $(10^4-10^8)$~sec is already
disfavored~\cite{Feng:2004mt}, while the stau NLSP is still a viable
scenario. The neutralino NLSP scenario can still be rescued if we
avoid the stringent BBN constraints, namely if the neutralino
lifetime becomes either larger than the age of the universe or lower
than the BBN time. The first possibility is realized in the
degenerate gravitino scenario~\cite{Boubekeur:2010nt}, where the
neutralino is extremely long-lived, and the only constraint comes
from the cold dark matter bound.

For neutralino NLSPs, the dominant decay mode is $\chi \to \gamma \gravitino$,
for which the decay width is~\cite{Feng:2004mt, Ellis:2003dn}
\begin{equation}
\Gamma(\chi \to \gamma \gravitino)
=
\frac{| N_{11} \cos \theta_W + N_{12} \sin \theta_W |^2}
{48\pi M_*^2} \
\frac{m_{\chi}^5}{m_{\gravitino}^2}
\left[1 - \frac{m_{\gravitino}^2}{m_{\chi}^2} \right]^3
\left[1 + 3 \frac{m_{\gravitino}^2}{m_{\chi}^2} \right]
\label{neutralinogamma}
\end{equation}
where $M_*$ is the Planck mass, $m_{\chi}$ is the neutralino mass,
and $\theta_W$ is the weak angle. This decay contributes only to EM
energy. If kinematically allowed, the neutralino will also decay to
gravitino and Z boson, or gravitino and light standard model Higgs
boson h. The leading contribution to hadronic energy is from $\chi
\to Z \gravitino,\, h \gravitino$.  These decays produce EM energy
for all possible $Z$ and $h$ decay modes (except $Z \to \nu
\bar{\nu}$), but they may also produce hadronic energy when followed
by $Z, h \to q \bar{q}$.  The decay width to $Z$ bosons
is~\cite{Feng:2004mt, Ellis:2003dn}
\begin{eqnarray}
\Gamma (\chi \to Z \gravitino) &=&
\frac{| - N_{11} \sin \theta_W + N_{12} \cos \theta_W |^2}
{48\pi M_*^2} \frac{\mchi^5}{\mgravitino^2}
F(\mchi,\mgravitino,m_Z) \nonumber \\
&& \times \left[ \left( 1-\frac{\mgravitino^2}{\mchi^2} \right)^2
\left( 1 + 3 \frac{\mgravitino^2}{\mchi^2} \right)
-\frac{m_Z^2}{\mchi^2}  G (\mchi,\mgravitino,m_Z) \right]
\end{eqnarray}
where
\begin{eqnarray}
F (\mchi,\mgravitino,m_Z) &=& \left[
\left(1- \left( \frac{\mgravitino + m_Z}{\mchi} \right)^2 \right)
\left(1- \left( \frac{\mgravitino - m_Z}{\mchi} \right)^2 \right)
\right]^{1/2} \label{F} \\
G (\mchi,\mgravitino,m_Z) &=& 3 + \frac{\mgravitino^3}{\mchi^3}
\left(-12 + \frac{\mgravitino}{\mchi} \right)
 + \frac{m_Z^4}{\mchi^4} - \frac{m_Z^2}{\mchi^2}
\left(3 - \frac{\mgravitino^2}{\mchi^2} \right)
\label{G}
\end{eqnarray}
with $m_Z \simeq 91$~GeV the mass of the Z boson.

The decay width to the Higgs boson is~\cite{Feng:2004mt, Ellis:2003dn}
\begin{eqnarray}
\Gamma (\chi \to h \gravitino) &=&
\frac{| N_{13} S_{11} + N_{14} S_{12} + N_{15} S_{13} |^2}
{48\pi M_*^2} \frac{\mchi^5}{\mgravitino^2}
F(\mchi,\mgravitino,m_h) \nonumber \\
&& \times \left[ \left( 1-\frac{\mgravitino}{\mchi} \right)^2
\left( 1 + \frac{\mgravitino}{\mchi} \right)^4
-\frac{m_h^2}{\mchi^2}  H (\mchi,\mgravitino,m_h) \right]
\label{hwidth}
\end{eqnarray}
where $S_{ij}$ are
three of the components of the mixing matrix in the
Higgs sector for the CP-even mass eigenstates, $F$ is
as given in \eqref{F}, and
\begin{eqnarray}
H(\mchi,\mgravitino,m_h) &=& 3 + 4\frac{\mgravitino}{\mchi}
+2 \frac{\mgravitino^2}{\mchi^2}+4 \frac{\mgravitino^3}{\mchi^3}
+3\frac{\mgravitino^4}{\mchi^4}
+ \frac{m_h^4}{\mchi^4} \nonumber \\
&& - \frac{m_h^2}{\mchi^2}
\left(3 +2 \frac{\mgravitino}{\mchi}
+3\frac{\mgravitino^2}{\mchi^2} \right)
\end{eqnarray}
with $m_h$ the mass of the Higgs boson.
Therefore, the neutralino lifetime is given by
\begin{eqnarray}
\tau & = & \frac{1}{\Gamma} \\
\Gamma & = & \Gamma(\chi \to \gamma \gravitino)+\Gamma(\chi \to Z
\gravitino)+\Gamma(\chi \to h \gravitino)
\end{eqnarray}
Given these two-body decay widths, the resulting values for the
energy release parameters are
\begin{eqnarray}
B_{\text{EM}}^{\chi} &\simeq& 1 \\
\epsilon_{\text{EM}}^{\chi} &=&
\frac{m_{\chi}^2 - m_{\gravitino}^2}{2 m_{\chi}} \\
B_{\text{had}}^{\chi} &\simeq&
\frac{\Gamma(\chi \to Z \gravitino) B_{\text{had}}^Z
+ \Gamma(\chi \to h \gravitino) B_{\text{had}}^h
+ \Gamma(\chi \to q \bar{q} \gravitino) }
{\Gamma(\chi \to \gamma \gravitino) +
\Gamma(\chi \to Z \gravitino) +
\Gamma(\chi \to h \gravitino)} \\
\epsilon_{\text{had}}^{\chi} &\approx&
\frac{m_{\chi}^2 - m_{\gravitino}^2 + m_{Z,h}^2}{2 m_{\chi}} \ ,
\end{eqnarray}
where $B_{\text{had}}^h \approx 0.9$,  $B_{\text{had}}^Z \approx 0.7$,
and the three-body
decay $\Gamma ( \chi \to q \bar{q} \gravitino) \sim 10^{-3} \: \Gamma$
~\cite{Feng:2004mt}.

- Finally, it must be noted that for long neutralino lifetimes,
$\tau_{NLSP} \geq 10^7$~sec, in addition to BBN constraints there
are strong bounds from the shape of the cosmic microwave background
(CMB) black-body spectrum~\cite{Boubekeur:2010nt, Ellis:1984eq}.
However, in our investigation we have found that the neutralino
lifetime is always $\tau_{NLSP} \leq 10^2$~sec, for gravitino masses
$\mgravitino \leq 1$~GeV, and therefore we do not need to worry
about these bounds from the CMB shape. 

Our main results are summarized in the figures below. Before starting
to discuss the figures, let us first make a few comments. The precise neutralino
composition depends on the values of the coefficients $N_{1i}$ in
(\ref{composition}), which in turn depend on the values of the free parameters
of the model. Roughly, for large coupling $\lambda=0.1-0.5$ the neutralino is
mainly a bino, while for small coupling $\lambda \ll 1$, the neutralino is mostly
a singlino. We refer the interested reader to e.g.~\cite{Hugonie:2007vd} for
the relevant discussion. Furthermore, the neutralino lifetime is determined
by the three decay channels to gravitino plus photon or Z boson or Higgs boson.
These partial decay rates depend on the available phase space (masses) as well as
the couplings (composition coefficients $N_{1i}$). Thus in the bino case, in
which $N_{11} \simeq 1$ and the rest of the coefficients are very small, the decay
rate to gravitino and Higgs boson is negligible, while in the singlino case,
in which $N_{15} \simeq 1$ and the rest of the coefficients are very small,
the decay rate to gravitino and Z boson is negligible. The decay channel to
gravitino and photon gives the main contribution, while the decay rate to
gravitino and Z boson (in the bino case) or to gravitino and Higgs boson (in the
singlino case) modify the neutralino lifetime by a factor of twenty or fifty
per cent respectively. Finally, for a given point in the cNMSSM parameter space, the
neutralino lifetime
is a function of the gravitino mass only. Imposing the BBN constraints we find
the maximum allowed
gravitino mass, and from the cold dark matter bound we can determine
the maximum allowed reheating temperature.

We can now turn to the figures where we discuss the two cases (bino or singlino)
separately. The first three figures correspond to the bino case, while the last
two figures correspond to the singlino case. For the bino case, it is important
to notice that the neutralino relic density can take values larger than the usual
ones by two
orders of magnitude. The reheating temperature decreases
with the neutralino relic density, and takes larger values for very low neutralino
relic density. Figure~3 shows the maximum allowed reheating temperature after
inflation versus the maximum allowed gravitino mass, both in GeV. Although it
cannot be seen directly from the figures, the maximum possible gravitino mass
in the bino case is $\mgravitino \simeq 1$~GeV, and the corresponding
reheating temperature is $T_R \sim 10^7$~GeV. Therefore, we see that a) the gravitino
in this scenario must be much lighter than the rest of superpatners, and b) the reheating
temperature after inflation is not large enough for thermal leptogenesis.
We remark in passing that gravitino
masses of the order of $1$~GeV can be obtained in
hybrid supersymmetric models of gauge- and gravity-mediation, in
which gravity provides sub-dominant and yet non-negligible
contributions~\cite{Hiller:2008sv}.

For the singlino case, we show in figure~4, the gravitino mass in
GeV versus neutralino relic density, and in figure~5 the maximum
allowed reheating temperature versus gravitino mass, both in GeV.
This time the neutralino relic density is even larger than before,
and gravitino now must be extremely light. This is due to the
smallness of the coefficients $N_{11}, N_{12}$ in the decay rate to
gravitino and photon. For the same lifetime as before, the gravitino
mass must be several orders of magnitude lower than in the bino
case. The last figure shows that in the singlino case the reheating
temperature cannot be larger than about $200$~GeV. However, this
value is much lower than the minimum value required for the
computation of the gravitino thermal production, and therefore we
conclude that this scenario must be excluded. We can understand
these features as follows. First, recall that the WMAP bound for the
cold dark matter abundance relates the gravitino mass to the
reheating temperature, and we can obtain an upper bound on the
reheating temperature for a given gravitino mass. From equation
(27), and for a gluino mass $m_{\tilde{g}} \sim 1$~TeV, we see that
when the gravitino is extremely light, $\mgravitino \simeq
10^{-6}$~GeV, the upper bound on the reheating temperature becomes
$T_R \simeq 41$~GeV. We then need to understand why the gravitino becomes
so light in the singlino case. Let us assume that we have in the
parameter space a point that corresponds to the bino case, another
point that corresponds to the singlino case, and that the Higgs
mass, superpartner masses, as well as the neutralino abundance are
the same for the two points. The only thing that is different is the
composition coefficients for the neutralino. In the bino case the
first coefficient is practically unity and the rest tiny, while in
the singlino case the last coefficient is almost unity and the rest
negligible. The BBN constraints determine the maximum possible
gravitino lifetime, which is given essentially by the photon decay
channel. We thus have for the neutralino lifetime
\begin{equation}
\tau \sim \frac{\mgravitino^2}{|N_{11}|^2}
\end{equation}
Therefore if in the singlino case $|N_{11}| \simeq 10^{-6}$, the
gravitino mass becomes as low as $\mgravitino \simeq 10^{-6}$~GeV.

\section{Conclusions}

In the framework of the cNMSSM, we have considered a possible cosmological scenario
with the gravitino LSP and the neutralino NLSP. The gravitino is
stable and plays the role of cold dark matter in the universe, while
the neutralino is unstable and decays to gravitino. We
have taken into account the relevant gravitino production mechanisms,
which are i) the NLSP decay, and ii)
scattering processes from the thermal bath. Our results can be
seen in the figures. We have found that i) the gravitino
is necessarily very light, and ii) the reheating temperature after inflation
is two orders of magnitude lower than the temperature required for thermal
leptogenesis. The singlino scenario must be excluded, while in the bino case
it is possible to have a gravitino in the gravity-mediated SUSY-breaking scheme.

\section*{Acknowledgments}

It is a pleasure to thank O.~Vives, L.~Boubekeur, and
R.~N.~Hodgkinson for discussions. The authors acknowledge financial
support from spanish MEC and FEDER (EC) under grant FPA2008-02878,
and Generalitat Valenciana under the grant PROMETEO/2008/004.


\newpage


\begin{figure}
\centerline{\epsfig{figure=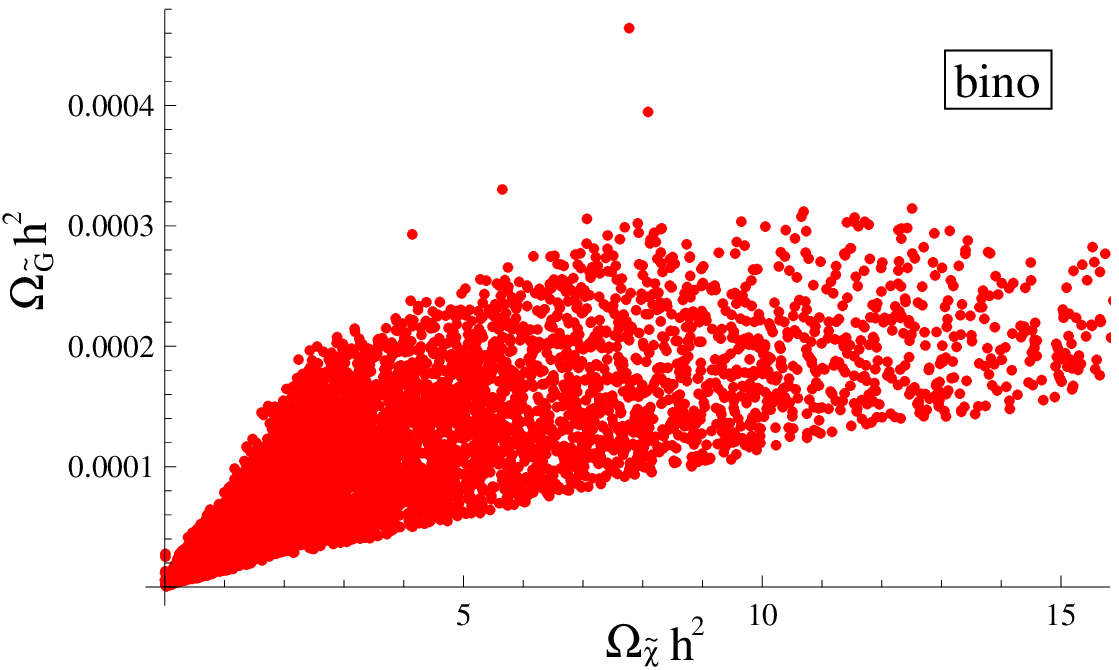,height=7cm,angle=0}}
\caption{Gravitino non-thermal production versus neutralino relic density for
the bino case.}
\end{figure}

\begin{figure}
\centerline{\epsfig{figure=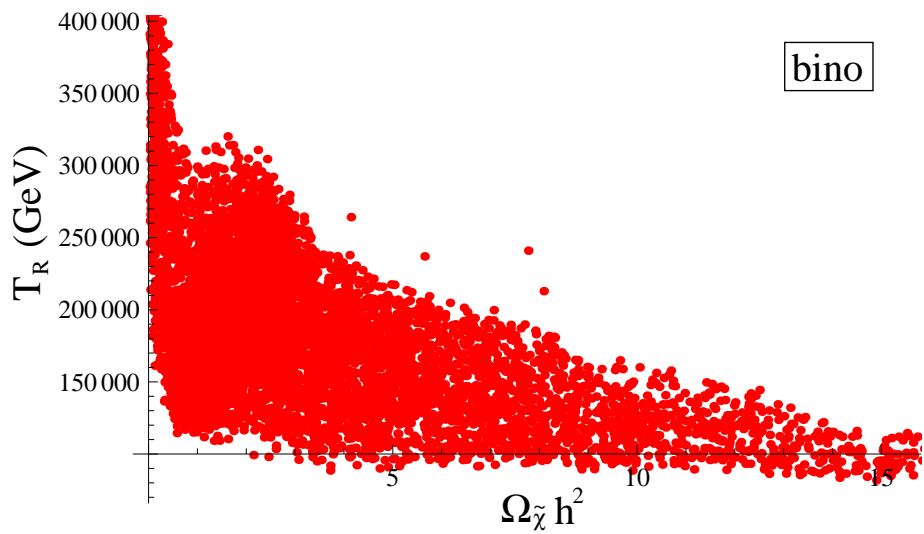,height=7cm,angle=0}}
\caption{Reheating temperature (in GeV) versus neutralino relic density
for the bino case.}
\end{figure}

\begin{figure}
\centerline{\epsfig{figure=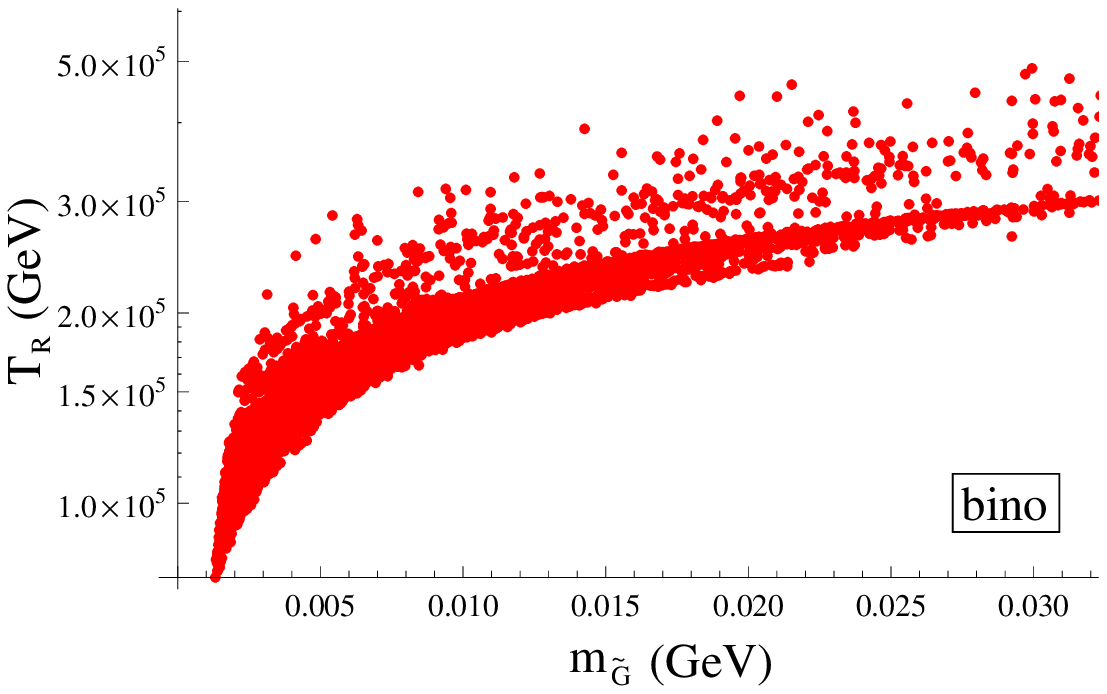,height=7cm,angle=0}}
\caption{Reheating temperature versus gravitino mass (both in GeV) for
the bino case.}
\end{figure}

\begin{figure}
\centerline{\epsfig{figure=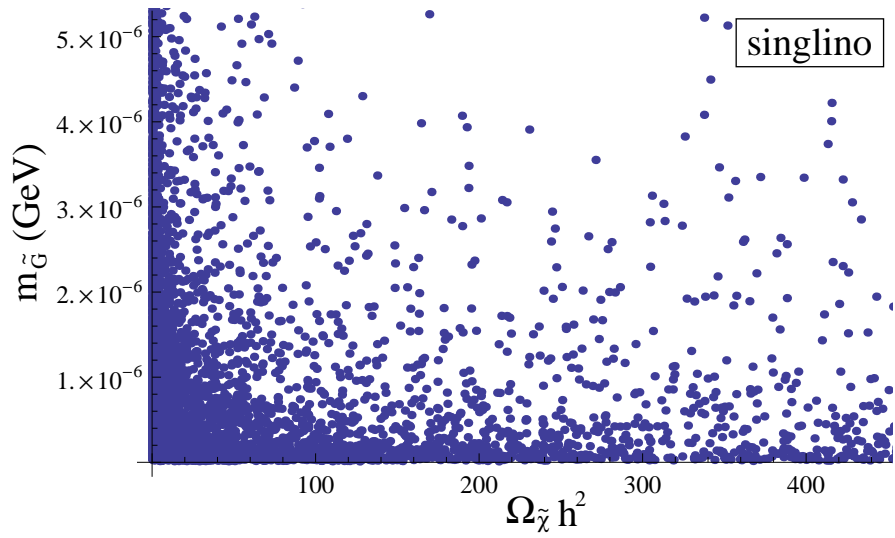,height=7cm,angle=0}}
\caption{Gravitino mass (in GeV) versus neutralino relic density for
the singlino case.}
\end{figure}

\begin{figure}
\centerline{\epsfig{figure=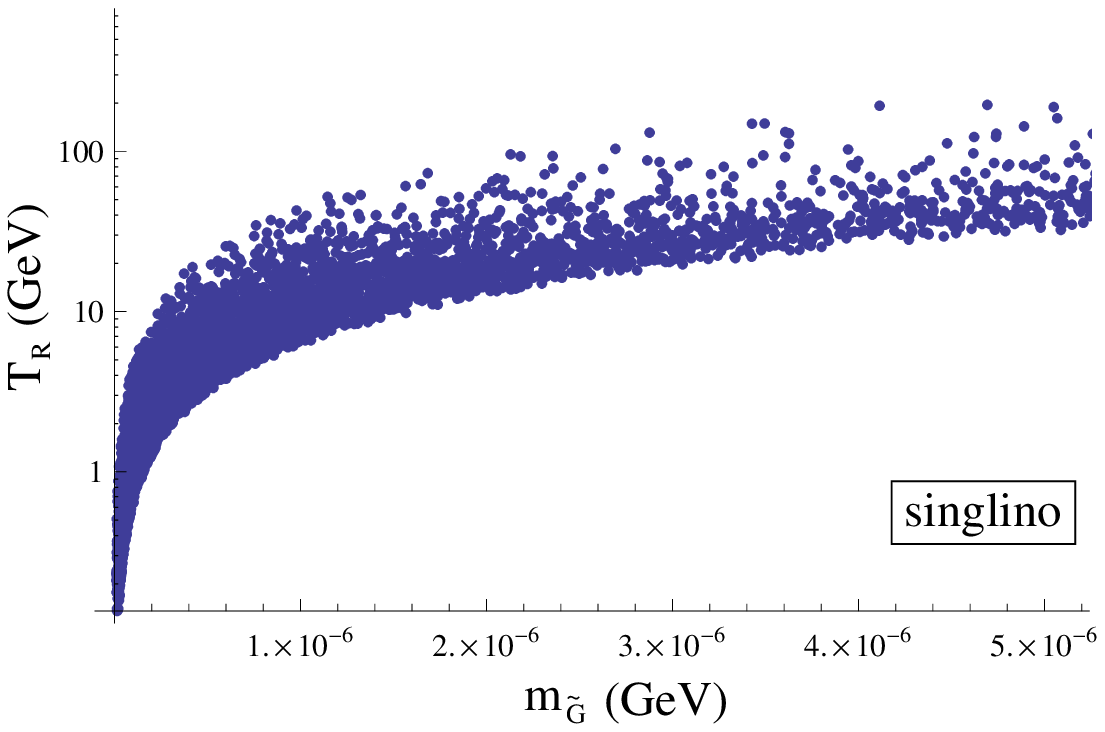,height=7cm,angle=0}}
\caption{Reheating temperature versus gravitino mass (both in GeV) for
the singlino case.}
\end{figure}

\end{document}